\documentclass[aps,pra,twocolumn,superscriptaddress]{revtex4}

\usepackage{amsmath,amsfonts,amssymb}
\usepackage{graphicx,float,calc}
\usepackage{color,bm}
\usepackage{ulem}
\usepackage{braket}
\usepackage{hyperref}
\usepackage{graphicx,epstopdf}
\usepackage{CJK}

\newcommand{\ehbar}{\hbar_{\text{eff}}}

\begin{document}
\title{Scaling of out-of-time ordered correlators in a non-Hermitian kicked rotor model}

\author{Wen-Lei Zhao}
\email{wlzhao@jxust.edu.cn}
\affiliation{School of Science, Jiangxi University of Science and Technology, Ganzhou 341000, China}
\author{Ru-Ru Wang}
\affiliation{School of Science, Jiangxi University of Science and Technology, Ganzhou 341000, China}

\begin{abstract}
We investigate the dynamics of the out-of-time-ordered correlators (OTOCs) via a non-Hermitian extension of the quantum kicked rotor model, where the kicking potential satisfies $\mathcal{PT}$-symmetry. The spontaneous $\cal{PT}$-symmetry breaking emerges when the strength of the imaginary part of the kicking potential exceeds a threshold value. We find, both analytically and numerically, that in the broken phase of $\cal{PT}$ symmetry, the OTOCs rapidly saturate with time evolution. Interestingly, the late-time saturation value scales as a pow-law in the system size. The mechanism of such scaling law results from the interplay between the effects of nonlocal operator in OTOCs and the time reversal induced by non-Hermitian driven potential.
\end{abstract}
\date{\today}

\maketitle
\section{Introduction}
In recent years, the out-of-time-ordered correlators (OTOCs) $C=-\langle [A(t),B]^2\rangle$ have attracted extensive attentions in diverse fields of physics, ranging from quantum chaos~\cite{Maldacena16,Rozenbaum17}, quantum information~\cite{Harris22} to black  holes physics~\cite{Hayden07}. A fundamental concept in these fields is the information scrambling, namely, the spread of information encoding in local degrees of freedoms over entire system so as to inaccessible by local measurement~\cite{Zanardi21,Touil21,Prakash20}. This progress is quantified by the growth of local operators with time evolution, due to which it will be no longer commutable with other operators separated long distance~\cite{TianciZhou20,SKZhao22}. The operator growth is dominated by the classical chaos in a way the rate of exponential growth of OTOCs is proportional to the classical Lyapunov exponent~\cite{Yin21,Moudgalya19}. Nowadays, the OTOCs have been widely utilized to diagnose the many-body localization~\cite{RHFan17,Smith19}, quantum entanglement~\cite{Garttner18,Keyserlingk18,Lerose20}, quantum thermalization~\cite{Lewis-Swan19,Zhu22,Balachandran21}, and many-body chaos~\cite{Kobrin21,Borgonovi19,Shenglong22}, hence receiving intensive investigations in the field of many-body physics~\cite{Heyl2018}. Interestingly, experimental advances have observed both the quantum information scrambling and quantum phase transition by measuring the OTOCs in the system of quantum circuit~\cite{Mi21} and a nuclear magnetic resonance quantum simulator~\cite{Nie20}.

For $\cal{PT}$-symmetric systems, the dynamics of OTOCs signals the Yang-Lee edge singularity~\cite{LJZhai20} of phase transition and displays the quantized response to external driven potential~\cite{WlZhao22}. It is now widely accepted that the non-Hermiticity is a fundamental modification to conventional quantum mechanics~\cite{Berry2004,Ashida20,Bender1998,Bender2002,Zhaoxm2021,Yu2021,Zhaoxm20211}, since many systems such as optics propagation in ``gain-or-loss'' medium~\cite{Javier22,Ganainy18}, the electronics transport in dissipative circuit~\cite{Xiao19,Chitsazi17,Deyuan21}, and cold atoms in tailored magneto-optical trap~\cite{Kreibich16,Keller97,Li19,YongmeiXue,Zejian22}, are described by non-Hermitian theory. The extension of Floquet systems to non-Hermitian regime even unveils rich physics~\cite{Longwen22,Longwen21,Longwen21B,DaJian19,Longhi17}. For example, the scaling of the spontaneous $\cal{PT}$-symmetry breaking and its relation with classical chaos are uncovered in a non-Hermitian chaotic system~\cite{West10}. The ballistic energy diffusion~\cite{Longhi2017} and quantized acceleration of momentum current~\cite{Zhao19} are reported in a $\cal{PT}$-symmetric kicked rotor (PTKR) model. The quantum kicked rotor (QKR) and its variants provide ideal platforms for investigating fundamental problems, such as the quantum transport in momentum-space lattice~\cite{Santhanam22,Ho12}, the quantum-to-classical transition of chaotic systems~\cite{Gadway13,Huang21}, as well as quantum thermalization in many-body systems~\cite{Vuatelet21}. The operator growth and chaotic information scrambling in different variations of QKR are still open issue and deserve urgent investigations.

In this context, we investigate, both analytically and numerically, the dynamics of OTOCs in a PTKR model, with focus on the broken phase of $\cal{PT}$ symmetry. We find that the OTOCs rapidly saturate with time evolution. Interestingly, the saturation value is the power-law function of the dimension of the system, which demonstrates a kind of scaling-law of the OTOCs with the system size. The mechanism of such scaling law results from two aspects. One is that the action of a nonlocal operators constructing the OTOCs on the state leads to a power-law decayed distribution in momentum space. The another is that the non-Hermitian kicking potential induces the perfect time reversal of the quantum state in momentum space. By utilizing the power-law decayed quantum state, we analytically obtain the scaling of OTOCs with the size of momentum space, for which the OTOCs is the power-law function of the dimension of the system. This demonstrates the OTOCs unboundedly increases with the system size, revealing a kind of fast scrambling~\cite{Belyansky20,Kuwahara21}. Our result sheds light on the Floquet engineering of the fast scramblers in non-Hermitian map systems.

The paper is organized as follows. In Sec.~\ref{Sec-MResults}, we show our model and the scaling-law of OTOCs. In Sec.~\ref{Sec-analysis}, we present the theoretical analysis of the scaling law. Sec.~\ref{Sec-conc} contains the conclusion and discussion.

\section{model and results}\label{Sec-MResults}

The Hamiltonian of a PTKR reads
\begin{equation}\label{PTKRHamil}
H = \frac{p^2}{2} + V_K(\theta)\sum_{j=0}^\infty \delta(t-t_n) \;,
\end{equation}
with the kicking potential
\begin{equation}\label{KPoten}
V_K(\theta)=  K\left[ \cos(\theta) + i \lambda \sin(\theta)\right] \;,
\end{equation}
which satisfies the $\cal{PT}$ symmetry $V_K(\theta)=V_K^*(-\theta)$. Here $p=-i\hbar_{\rm{eff}}\partial/\partial\theta$ is the angular momentum operator and $\theta$ is the angle coordinate, which obeys the communication relation $[\theta,p]=i\ehbar$ with $\ehbar$ the effective Planck constant. The parameters $K$ and $\lambda$ control the strength of the real and imaginary part of the kicking potential, respectively. The time $t_n$ is integer, i.e., $t_n=1,2\ldots$, indicating the kicking number. All variables are properly scaled and thus in dimensionless units.
The eigenequation of angular momentum operator is $p|\phi_{n}\rangle=p_{n}|\phi_{n}\rangle$ with eigenstate $\langle \theta |\phi_{n}\rangle=e^{i n \theta}/\sqrt{2\pi}$ and eigenvalue $p_{n}=n\hbar_{\rm{eff}}$. On the basis of $|\phi_{n}\rangle$, an arbitrary quantum state can be expanded as $|\psi\rangle=\sum_{n}\psi_{n}|\phi_{n}\rangle$. The evolution of the quantum state from $t_{j}$ to $t_{j+1}$ is given by $|\psi(t_{j+1})\rangle=U|\psi(t_{j})\rangle$, where the Floquet operator $U$ takes the form
\begin{equation}\label{EvolOpet}
U=\exp\left(-i\frac{p^2}{2\hbar_{\rm{eff}}}\right)\exp\left[
-i\frac{V_K(\theta)}{\hbar_{\rm{eff}}}\right]\;.
\end{equation}

The OTOCs are defined as the average of the squared commutator, i.e., $C(t)=-\langle[A(t),B]^2\rangle$, where both operators $A(t)=U^{\dagger}(t)AU(t)$ and $B$ are evaluated in the Heisenberg picture, and the $\langle \cdots\rangle=\langle \psi(t_0)|\cdots|\psi(t_0)\rangle$ indicates the expectation value taken over the initial state $|\psi(t_0)\rangle$~\cite{Heyl2018}. It usually uses the thermal states for taking the average in investigation of OTOCs of lattice systems. For the Floquet-driven system, however, there is no the well-defined thermal states, as the temperature trends to be infinity as time evolves~\cite{DAlessio14}. Without loss of generality, we choose a Gaussian wavepacket as an initial state, i.e., $\psi{(\theta,0)}=(\sigma/\pi)^{1/4} \exp (-\sigma \theta^{2}/2)$ with $\sigma=10$. We consider the case with $A=\theta$ and $B=p^{m}$ ($m\in\mathbb{N}$), hence $C(t)=-\langle[\theta(t_{}),p^{m}]^2\rangle$.

\begin{figure}[t]
\begin{center}
\includegraphics[width=8.0cm]{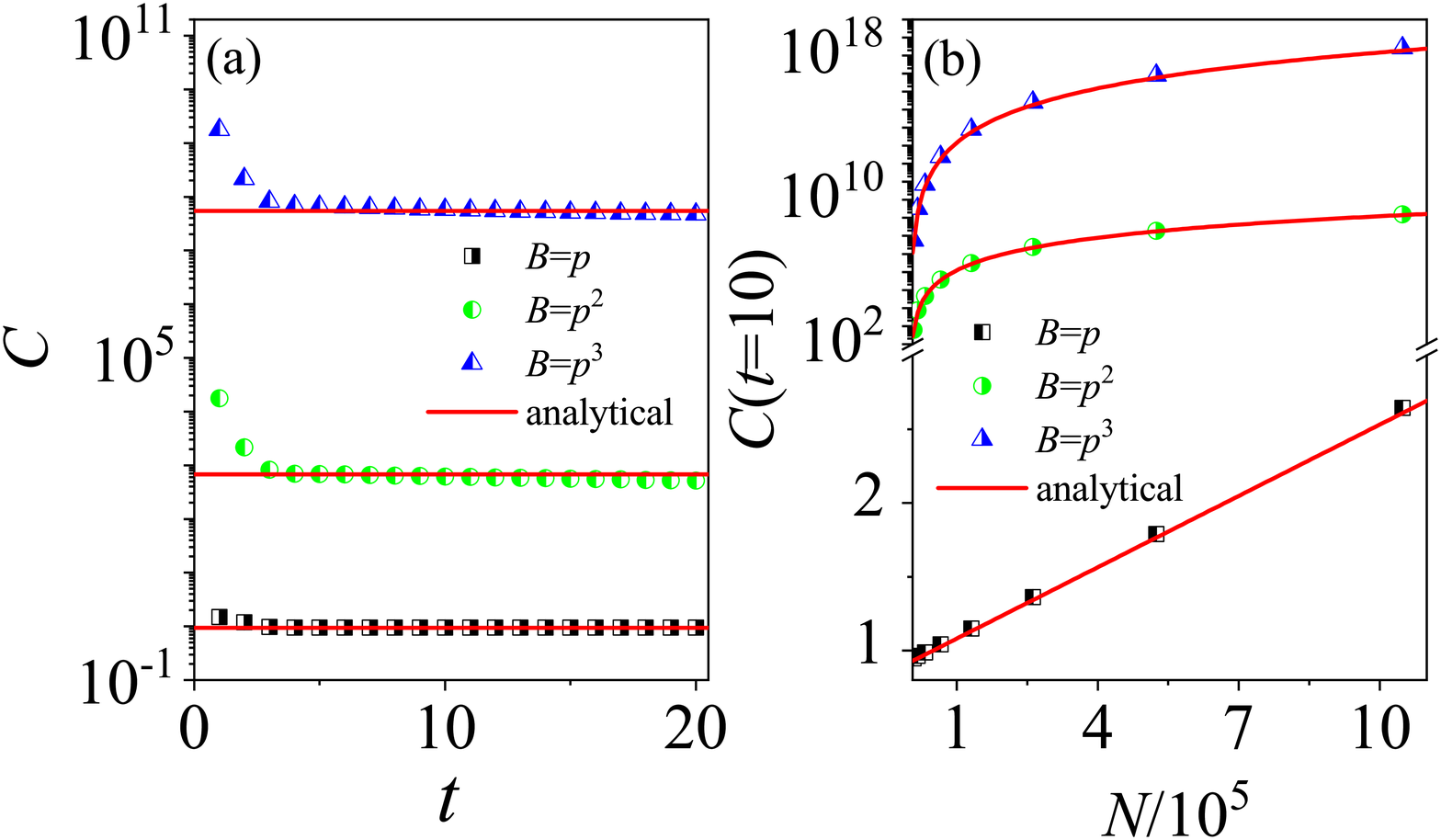}
\caption{(a) Time dependence of $C(t)$ with $B=p$ (squares), $p^{2}$ (circles), and $p^{3}$ (triangles) with $N=2^{13}$. (b) The $C(t)$ at the time $t=t_{10}$ versus $N$. Solid lines in (a) and (b) denote our theoretical prediction in Eq.(2). The parameters are $K=2\pi$, $\lambda=0.9$, and $\hbar_{\rm{eff}}=0.1$.}\label{OTOCSCLaw}
\end{center}
\end{figure}

Our main result is the scaling law of the OTOCs
\begin{equation}\label{EqScalling}
  C{(t)}\sim N^{2m-1} \theta^{2}_{c}\;,
\end{equation}
where $N$ is the dimension of the momentum space of the PTKR model, and $\theta_c=\pi/2$.
The above prediction is verified by numerical results in Fig.~\ref{OTOCSCLaw}. As an illustration, we consider $m=1$, 2, and 3 in numerical simulations. Figure~\ref{OTOCSCLaw}(a) shows that for a specific $m$, the $C(t)$ saturates rapidly as time evolves, which is in perfect agreement with our theoretical prediction in Eq.~\eqref{EqScalling}. In order to further confirm the scaling law of $C(t)$, we numerically investigate the $C$ at a specific time for different $N$. Figure~\ref{OTOCSCLaw}(b) shows that for $t=t_{10}$, the value of $C$ increases in the power-law of $N$, which coincides with the theoretical prediction in Eq.~\eqref{EqScalling}.  The scaling of $C(t_{})$ with dimension of the system demonstrates that it diverges as $N\rightarrow\infty$, which is of interest in the study of fast scrambling~\cite{Kuwahara21}. We would like to mention that we previously found the scaling law for the OTOCs constructed by $A=\theta$ and $B=p$, in a Gross-Pitaevski map system~\cite{Zhao21}. Our present work explores the scaling law for $B=p^m$ with arbitrary integer $m$, moreover extends the investigation to non-Hermitian systems, which is obviously an significant advance in the fields of operator growth in chaotic systems.
%%%%

\section{THEORETICAL ANALYSIS}\label{Sec-analysis}

\subsection{Analysis of the $C_1(t)$}

Straightforward derivation yields the expression of OTOCs
\begin{equation}\label{Otcexpansion}
  C{(t_{})}=C_{1}(t_{})+C_{2}(t_{})-2\text{Re}[C_{3}(t_{})]\;,
\end{equation}
where the three terms in right side are defined by
\begin{equation}\label{C1Defini}
  C_{1}(t_{})=\langle\psi_{R}(t_{0})|p^{2m}|\psi_{R}(t_{0})\rangle\;,
\end{equation}
\begin{equation}\label{C2Defini}
  C_{2}(t_{})=\langle\varphi_{R}(t_{0})|\varphi_{R}(t_{0})\rangle\;,
\end{equation}
and
\begin{equation}\label{C3Defini}
  C_{3}(t_{})=\langle\psi_{R}(t_{0})|p^m|\varphi_{R}(t_{0})\rangle\;,
\end{equation}
with $|\psi_R(t_0)\rangle=U^\dag(t_n,t_0)\theta U(t_n,t_0)|\psi(t_0)\rangle$ and $|\varphi_R(t_0)\rangle=U^\dag(t_n,t_0)\theta U(t_n,t_0)p^m|\psi(t_0)\rangle$.

To get the state $|\psi_R(t_0)\rangle$, one needs three steps: i). forward evolution $t_0 \rightarrow t_n$, i.e., $|\psi(t_n)\rangle=U(t_0,t_n)|\psi(t_0)$, ii). action of the operator $\theta$ on $|\psi(t_n)\rangle$, i.e., $|\tilde{\psi}(t_n)\rangle= p |\psi(t_n)\rangle$, and iii). backward evolution $t_n \rightarrow t_0$, i.e., $|\psi_R(t_0)\rangle={U}^{\dagger}(t_0,t_n)|\tilde{\psi}(t_n)\rangle$. The $C_1(t_n)$ [see Eq.~\eqref{C1Defini}] is just expectation value of the $p^{2m}$ taken over the state $|\psi_R(t_0)\rangle$. For the numerical calculation of the state $|\varphi_R(t_0)\rangle$, one should first construct the operation of $p^{m}$ on the initial state $|\psi(t_0)\rangle$, i.e., $|\varphi(t_0)\rangle=p|\psi(t_0)\rangle$. Then, forward evolution from $t_0$ to $t_n$ yields the state $|\varphi(t_n)\rangle=U(t_0,t_n)|\varphi(t_0)\rangle$. At the time $t=t_n$, the action of $p^{m}$ on the state $|\varphi(t_n)\rangle$ results in a new state $|\tilde{\varphi}(t_n)\rangle=p|\varphi(t_n)\rangle$, starting from which the time reversal process $t_n \rightarrow t_0$ yields the state $|\varphi_R(t_0)\rangle={U}^{\dagger}(t_0,t_n)|\varphi(t_0)\rangle$.
The norm of the $|\varphi_R(t_0)\rangle$ is just the $C_2(t_n)$ [see Eq.~\eqref{C2Defini}]. As the two states $|\psi_R(t_0)\rangle$ and $|\varphi_R(t_0)\rangle$ are available at the end of time reversal, one can calculate the $C_3(t_n)$ according to Eq.~\eqref{C3Defini}].

It is known that in the $\cal{PT}$-symmetry breaking phase, the norm of quantum state $\mathcal{N}_{\psi}(t_n)=\langle\psi(t_n) | \psi(t_n) \rangle$ exponentially increases for both the forward and backward time evolution. To eliminate the contribution of norm to OTOCs, it is necessary to take the normalization for the time-evolved state.  Specifically, for the forward evolution $t_0\rightarrow t_n$, we set the norm of the quantum state equals to that of the initial state, i.e., $\mathcal{N}_{\psi}(t_j)=\langle\psi(t_0) | \psi(t_0) \rangle$ with $0\leq j \leq n$. The backward evolution starts from the time $t=t_n$ with the state $|\tilde{\psi}(t_n)\rangle$, whose norm $\mathcal{N}_{\tilde{\psi}}(t_n)=\langle\psi(t_n)|\theta^2| \psi(t_n) \rangle$ is expectation value of $\theta^{2}$ with the state $| \psi(t_n) \rangle$. It is apparent that the value of $\mathcal{N}_{\tilde{\psi}}(t_n)$ is an important information encoded by the operation of $\theta$ on the state $| \psi(t_n) \rangle$. Based on this, we take the normalization of the quantum state in the backward evolution $t_n\rightarrow t_0$ in such a way that its norm equals to $\mathcal{N}_{\tilde{\psi}}(t_n)$, i.e., $\mathcal{N}_{\psi_R}(t_j)=\mathcal{N}_{\tilde{\psi}}(t_n)$. One can find that for both the forward and backward evolution, the norm of a time-evolved state always equals to that of the state which the time evolution starts from. The the same procedure of normalization is applied in calculating $C_2(t_n)$. Therefore, we have the equivalence $\mathcal{N}_{\varphi}(t_j)=\langle \varphi(t_0)|\varphi(t_0)\rangle$ and $\mathcal{N}_{\varphi_R}(t_j)=\langle \tilde{\varphi}(t_n)|\tilde{\varphi}(t_n)\rangle$ ($0\leq j \leq n$) for the forward evolution and time reversal, respectively.
%%%%
\begin{figure}[htbp]
\begin{center}
\includegraphics[width=8cm]{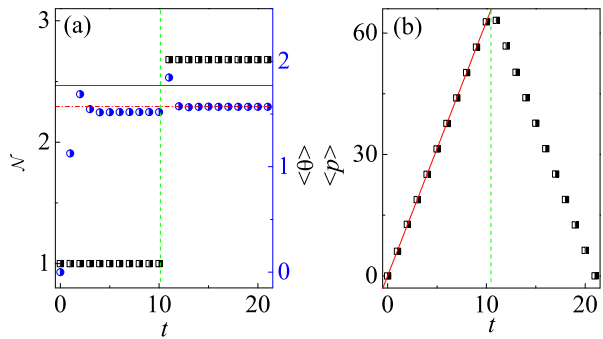}
\caption{Time evolution of $\mathcal{N}$ (a), $\langle\theta\rangle$ (a), and $\langle p \rangle$ (b) with $t=t_{10}$. In (a): Solid and dash-dotted lines indicate $\mathcal{N}=\theta^2_c$ and $\langle\theta\rangle=\theta_c (=\pi/2)$, respectively. In (b): Solid line indicates $\langle p \rangle=K t$. Green dashed lines in (a) and (b) are auxiliary lines. The parameters are the same as in Fig.~\ref{OTOCSCLaw}.}\label{ObsTReversal}
\end{center}
\end{figure}

\begin{figure}[t]
\begin{center}
\includegraphics[width=8.0cm]{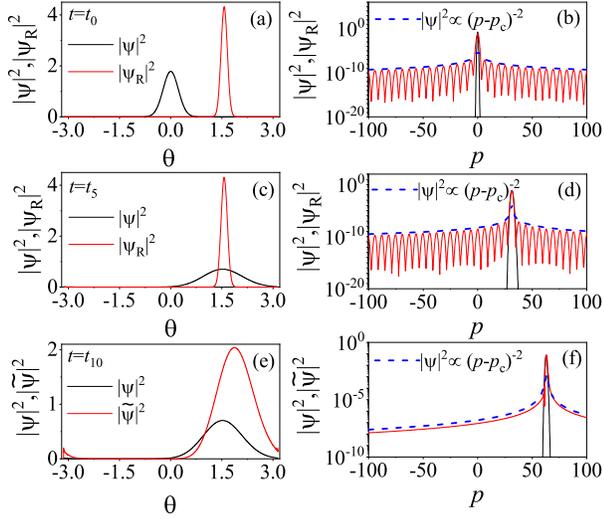}
\caption{Distributions in real (left panels) and momentum (right panels) space. In (a)-(d): Black and red lines indicate the distribution of the states at the forward $|\psi(t_j)\rangle$ and backward $|\psi_R(t_j)\rangle$ evolution, respectively, with $t=t_{0}$ (top panels), $t=t_5$ (middle panels), and $t=t_{10}$ (bottom panels). In (e)-(f): Red and black lines indicate the distribution of the states $|\psi(t_{10})\rangle$ and $|\tilde{\psi}(t_{10})\rangle=\theta|\psi(t_{10})\rangle$. Blue dashed lines indicate the power-law decay $|\psi|^2 \propto (p-p_c)^{-2}$. The parameters are the same as in Fig.~\ref{OTOCSCLaw}.}\label{Distributions}
\end{center}
\end{figure}

We rewrite the $C_1$ as
\begin{equation}\label{C1Defini-II}
  C_{1}(t_{})=\langle\psi_{R}(t_{0})|p^{2m}|\psi_{R}(t_{0})\rangle=\langle p^{2m}(t_0)\rangle_R\mathcal{N}_{\psi_R}(t_0)\;,
\end{equation}
where $\mathcal{N}_{\psi_R}(t_0)=\langle \psi_R(t_0)|\psi_R(t_0)\rangle$ is the norm of the quantum state $|\psi_R(t_0)\rangle$ and $\langle p^{2m}(t_0)\rangle_R=\langle \psi_R(t_0)|{p}^{2m}|\psi_R(t_0)\rangle/\mathcal{N}_{\psi_R}(t_0)$ indicates the exception value of ${p}^{2m}$ of the state $|\psi_R(t_0)\rangle$ with the division of its norm.
We numerically investigate both the forward and backward evolution of the norm $\mathcal{N}$, and the mean values $\langle \theta\rangle$ and $\langle p\rangle$ for a specific time, e.g., $t=t_{10}$. Note that we define the expectation value of observable $Q$ as $\langle Q \rangle =\langle\psi(t)|Q|\psi(t)\rangle/\mathcal{N}(t)$ with $\mathcal{N}(t)=\langle\psi(t)|\psi(t)\rangle$. It is apparently that such kind of definition eliminates the contribution of norm to mean value. Figure~\ref{ObsTReversal}(a) shows that the norm is unity during the forward time evolution (i.e., $t_0\rightarrow t_{10})$, and remains at a fixed value, i.e., i.e., $\mathcal{N}_{\psi_R}(t_0)\approx \theta^2_c$ during the backward evolution (i.e., $t_{10} \rightarrow t_0$). For $t_0\rightarrow t_{10}$, the value of norm equals to that of the normalized initial state, so $\mathcal{N}(t_j)=1$. For the time reversal $t_{10}\rightarrow t_{0}$, our normalization procedure results in the equivalence $\mathcal{N}_{\psi_R}(t_j)=\langle \psi(t_n)|\theta^2|\psi(t_n)\rangle$. Interestingly, our numerical investigations in Figs.~\ref{Distributions}(a), (c) and (e) show that for the forward evolution, the initially Gaussian wavepacket rapidly moves to the position $\theta_c=\pi/2$. During the time reversal, it remains localized at $\theta_c$ with the width of distribution being much smaller than that of the state of forward evolution. Correspondingly, the mean value $\langle \theta\rangle$ has very slight differences with $\theta_c$ [see Fig.~\ref{ObsTReversal}(a)]. Since the quantum state is extremely localized at $\theta_c$, one can get the approximation
\begin{equation}\label{NrmPsiRT0}
  \mathcal{N}_{\psi_R}(t_0)=\langle\psi(t_n)|\theta^2|\psi(t_n)\rangle\approx \theta^2_c\;.
\end{equation}
%%%%%

Figures~\ref{Distributions}(b), (d) and (f) show the momentum distribution of the state during both forward and backward evolution. For the forward evolution, the quantum state behaves like a soliton which moves to positive direction in momentum space, resulting in the linear increase of mean momentum, i.e., $\langle p \rangle = Kt$ [see Fig.~\ref{ObsTReversal}(b)]. The mechanism of the directed acceleration has been unveiled in our previous investigation in Ref.~\cite{Zhao19,WlZhao22}. Intriguingly, at the time $t=t_{10}$, the action of $\theta$ yields a state with a power-decayed shape, i.e., $|\psi_R(p,t_0)|^2\propto (p-p_c)^{-2}$ [see Fig.~\ref{Distributions}(f)]. Most importantly, during backward evolution, the quantum state still retains the power-law decayed shape, for which the center $p_c$ decreases with time and almost overlaps with that of the state of forward evolution. This clearly demonstrates a kind of time reversal of transport behavior in momentum space.

\begin{figure}[t]
\begin{center}
\includegraphics[width=7.0cm]{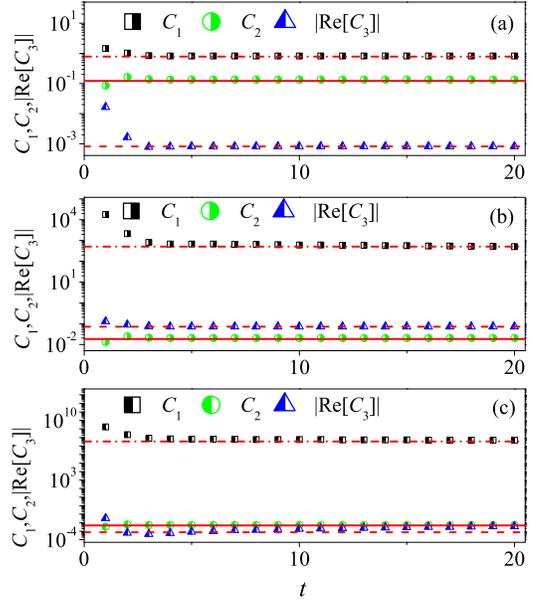}
\caption{The $C_1$ (squares), $C_2$ (circles), and $\left| \text{Re}[C_3]\right|$ (triangles) versus time with $B=p$ (a), $p^2$ (b), and $p^3$ (c). Dash-dotted, solid, and dashed lines indicate our theoretical prediction in Eq.~\eqref{C1Defini-III} for $C_1$, Eq.~\eqref{C2Defini-III} for $C_2$, and Eq.~\eqref{C3Defini-III} for $C_3$, respectively. The parameters are the same as in Fig.~\ref{OTOCSCLaw}.}\label{Threeparts}
\end{center}
\end{figure}

In the aspect of the mean momentum $\langle p \rangle$, we find that the value of $\langle p \rangle$ linearly decreases during the backward evolution and is in perfect symmetry with that of forward evolution, which is a solid evidence of time reversal. In the end of the backward evolution, the quantum state $|\psi_R(t_0)\rangle$ is localized at the point $p=0$ [see Fig.~\ref{Distributions}(b)]. By using the power-law distribution $|\psi_R(p,t_0)|^2\sim p^{-2}$, it is straightforwardly to get the estimation of the expectation value of $p^{2m}$, i.e., $\langle p^{2m}\rangle_{\psi_R}=\int_{p_{-N/2}}^{p_{N/2}}{p^{2m}|\psi_R(p,t_0)|^2 dp}\propto N^{2m-1}$. Taking both the $\langle p^{2m}\rangle_{\psi_R}$ and $\mathcal{N}_{\psi_R}(t_0)$ in Eq.~\eqref{NrmPsiRT0} into Eq.~\eqref{C1Defini-II} yields the relation
\begin{equation}\label{C1Defini-III}
C_1(t)\propto N^{2m-1} \theta_c^2\;,
\end{equation}
which is verified by our numerical results in Fig.~\ref{Threeparts}.
As an illustration, we consider the cases with $m=1$, 2 and 3. Our numerical results of the time dependence of $C_1$ is in good agreement with Eq.~\eqref{C1Defini-III}. It is now clear that the scaling of $C(t)$ with $N$ originates from the power-law decay of the state $|\psi_R(t_0)\rangle$. The reason for the formation of power-law decayed wavefuction has been uncovered in Ref.~\cite{Zhao21}.
%%%%%

\subsection{Analysis of $C_2(t)$}

\begin{figure}[t]
\begin{center}
\includegraphics[width=7.0cm]{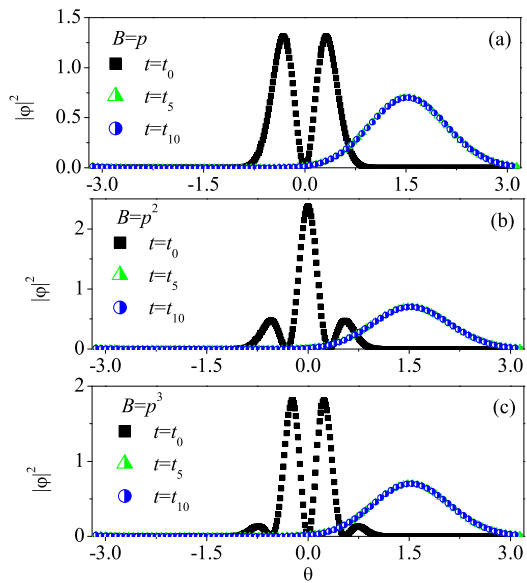}
\caption{ Probability density distributions in real space at the time $t=t_0$ (squares), $t_5$ (triangles), and $t_10$ (circles) with $B=p$ (a), $p^2$ (b), and $p^3$ (c). The parameters are the same as in Fig.~\ref{OTOCSCLaw}.}\label{C2Distris}
\end{center}
\end{figure}

We proceed to evaluate the time dependence of $C_2(t)$ in Eq.~\eqref{C2Defini}, which is just the norm of the state $|\varphi_{R}(t_{0})\rangle$ at the end of backward evolution. According to our normalization procedure, the value of $C_2$ equals to the norm of the state $|\tilde{\varphi}(t_n)\rangle=\theta |\varphi (t_n)\rangle$, hence
\begin{equation}\label{C2Defini-II}
C_2=\langle \varphi(t_n)|\theta^2|\varphi(t_n)\rangle=\langle \theta^2\rangle \mathcal{N}_{\varphi}(t_n)\;,
\end{equation}
with $\mathcal{N}_{\varphi}(t_n)=\langle \varphi(t_n)|\varphi(t_n)\rangle$ and $\langle \theta^2\rangle=\langle \varphi(t_n)|\theta^2|\varphi(t_n)\rangle/\mathcal{N}(t_n)$. We numerically find that the state $|\varphi(t_n)\rangle$ is extremely localized at the position $\theta_c$ during the forward evolution [see Fig.~\ref{C2Distris}]. Then a rough estimation yields $\langle \theta^2 \rangle\sim \theta_c^2$. The norm $\mathcal{N}_{\varphi}(t_n)$ equals to that of the initial state $|\varphi(t_0)\rangle=p^m|\psi(t_0)\rangle$. By using the initially Gaussian wavepacket $\psi(p,t_0)=(1/\sigma \ehbar^2 \pi)^{1/4}\exp(-p^2/2\sigma \ehbar^2)$, one can straightforwardly get
\begin{equation*}
\mathcal{N}_{\varphi}(t_n)=\int_{-\infty}^{\infty}p^{2m}|\psi(p,t_0)|^2dp=\frac{(2m-1)!!}{2^m
\alpha^m}\;,
\end{equation*}
where $\alpha=1/(\sigma\hbar^2)$ and $(...)!!$ denotes a double factorial. Taking both the $\langle \theta \rangle$ and $\mathcal{N}_{\varphi}(t_n)$ into Eq.~\eqref{C2Defini-II} yields
\begin{equation}\label{C2Defini-III}
  C_2(t)\sim \theta^2_c \frac{(2m-1)!!}{2^m \alpha^m}\;,
\end{equation}
which is in good agreement with our numerical results in Fig.~\ref{Threeparts}.
%%%%%%%%%%%

\subsection{Analysis of $C_3(t)$}

\begin{figure}[t]
\begin{center}
\includegraphics[width=8.5cm]{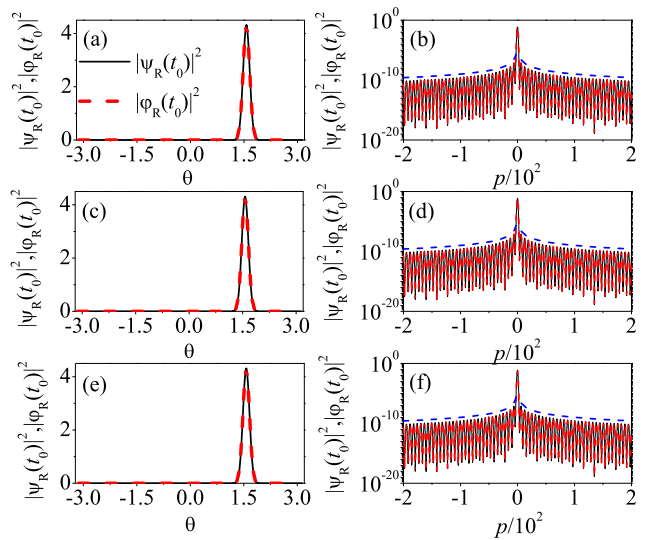}
\caption{Comparison of the distribution of states $|\psi _R(t_0)\rangle$ (solid lines) and $|\varphi _R(t_0)\rangle$ (dashed lines) in real (left panels) and momentum space (right panels) with $B=p$ (top panels), $p^2$ (middle panels), and $p^3$ (bottom panels). Blue dashed lines in (b), (d) and (f) indicate the power-law decay $|\psi_R(t_0)|^2(|\varphi _R(t_0)|^2) \propto p^{-2}$. The parameters are the same as in Fig.~\ref{OTOCSCLaw}.}\label{T0Distributions}
\end{center}
\end{figure}

The value of $C_3(t)$ depends on both the states $|\psi_R(t_0)\rangle$ and $|\varphi_R(t_0)\rangle$ [see Eq.~\eqref{C3Defini}]. Figure~\ref{T0Distributions} shows the probability density distributions of the two states in both the real space and momentum space. For comparison, the two states are normalized to unity. One can find the perfect consistence between $|\psi_R(t_0)\rangle$ and $|\varphi_R(t_0)\rangle$. Then, we roughly regard the $C_3$ as the expectation value of the $p^m$ taking over the state $\psi_R(t_0)$ or $\varphi_R(t_0)$, i.e., $C_3(t)\approx\langle p^m(t_0) \rangle_{\psi_R}\sqrt{\mathcal{N}_{\psi_R}(t_0)}\sqrt{\mathcal{N}_{\varphi_R}(t_0)}$, where according to above derivations $\mathcal{N}_{\psi_R}(t_0)=\theta^2_c$ and $\mathcal{N}_{\varphi_R}(t_0)=C_2(t)$. By using the power-law decayed wavepacket $|\psi_R(t_0)|^2\propto p^{-2}$, one can get the estimation
\begin{equation}
\begin{aligned}
  \langle p^m(t_0) \rangle_{\psi_R}&\approx\int_{p_{-N/2}}^{p_{N/2}}{p^{m}|\psi_R(p,t_0)|^2 dp}\\
  &\sim \begin{cases}
  0 & \text{for odd}\; m\;,\\
   N^{m-1} & \text{for even}\; m\;.
  \end{cases}
\end{aligned}
\end{equation}
Accordingly, the $C_3$ is approximated as
\begin{equation}\label{C3Defini-III}
  C_3(t)\sim \begin{cases}
  0 & \text{for odd}\; m\;,\\
  \eta N^{m-1} & \text{for even}\; m\;.
\end{cases}
\end{equation}
with the prefactor $\eta\propto\theta_c^2\left[{(2m-1)!!}/{2^m \alpha^m}\right]^{\frac{1}{2}}$.
%%%%%%%

\begin{figure}[t]
\begin{center}
\includegraphics[width=8.0cm]{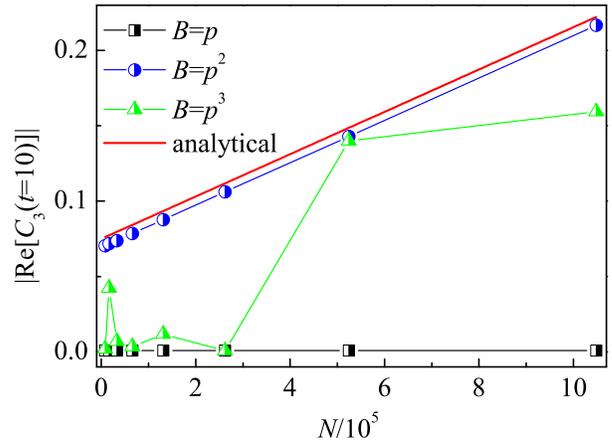}
\caption{The $|\text{Re}[C_3(t)]|$ at the time $t=t_{10}$ versus $N$ with $B=p$ (squares), $p^{2}$ (circles), and $p^{3}$ (triangles). Red solid line indicates our theoretical prediction in Eq.~\eqref{C3Defini-III} with $\eta=6.05\times10^{-7}$ for $B=p^2$. The parameters are the same as in Fig.~\ref{OTOCSCLaw}.}\label{C3Scal}
\end{center}
\end{figure}

We numerically calculate the absolute value of the real part of $C_3$. Interestingly, our numerical results of $|\text{Re}[C_3]|$ is in good agreement with the analytical prediction in Eq.~\eqref{C3Defini-III} [see Fig.~\ref{Threeparts}], which proves the validity of our theoretical analysis. We further numerically investigate the $|\text{Re}[C_3(t)]|$ at a specific time for different $N$. Figure~\ref{C3Scal} shows that for $B=p$, the value of $|\text{Re}[C_3(t)]|$ is nearly zero with varying $N$, which is consistent with our theoretical prediction in Eq.~\eqref{C3Defini-III}. For $B=p^3$, the value of $|\text{Re}[C_3(t)]|$ has slight difference with zero for large $N$, signaling the derivations with Eq.~\eqref{C3Defini-III}. This is due to the fact the quantum state $|\psi_R(t_0)|^2$ is not exactly symmetric around $p$. Interestingly, for $B=p^2$, the value of $|\text{Re}[C_3(t)]|$ increases linearly with increasing $N$, which is a clear evidence of the validity of our theoretical prediction.

\section{CONCLUSION AND DISCUSSION}\label{Sec-conc}

In the present work, we investigate the dynamics of the $C(t)=-\langle[\theta(t),p^m]\rangle$ in a PTKR model. The spontaneous $\cal{PT}$ symmetry breaking is assured by the condition $\lambda>\lambda_c$. In the broken phase of $\cal{PT}$-symmetry, we find, both analytically and numerically, the scaling law of $C(t)$ with the dimension of the momentum space, i.e., $C(t)\sim N^{2m-1}\theta_c^2$. This demonstrates that the value of $C$ increases unboundedly with $N$, which implies that the local perturbation can spread to the entire system very rapidly. In order to reveal the mechanism of the scaling, we make detailed investigations on both the forward and backward evolution of the quantum state. Our investigations show that the action of $\theta$ on a quantum state leads to the formation of the power-law decayed momentum distribution $|\psi(p)|^2\propto (p-p_c)^{-2}$. Interestingly, such a shape retains during the time reversal, besides the decrease of $p_c$ to almost zero. Based on the power-law decayed state, we analytically derive the time dependence of the three parts of the $C$, which is confirmed by numerical results.

In recent years, fruitful physics of quantum many-body systems, such as dynamical phase transition, many-body localization, and thermalization have received extensive studies. It is found that the energy conservation of chaotic systems leads to the scaling law of OTOCs, for which the late-time saturation of OTOCs scales as the inverse polynomial with the system size~\cite{Huang19}. For chaotic systems with long-range interaction, the time-dependence of OTOCs obeys the dynamical scaling law near the phase transition point~\cite{Wei19}. Accordingly, our finding of the power-law scaling of OTOCs with the system size of the PTKR model serves as a new element of the quantum information scrambling in non-Hermitian map systems.

\section*{ACKNOWLEDGMENTS}
Wen-Lei Zhao and Ru-Ru Wang are supported by the National Natural Science Foundation of China (Grant No. 12065009) and the Science and Technology Planning Project of Jiangxi province (Grant No. 20224ACB201006).


\begin{thebibliography}{10}

\bibitem{Maldacena16}
J. Maldacena, S. H. Shenker, and D. Stanford, A bound on
chaos, J. High Energy Phys. {\bf 08} 106 (2016).

\bibitem{Rozenbaum17}
E. B. Rozenbaum, S. Ganeshan, and V. Galitski, Lyapunov Exponent and Out-of-Time-Ordered Correlator's Growth Rate
in a Chaotic System, \prl {\bf 118}, 086801 (2017).

\bibitem{Harris22}
J. Harris, B. Yan, and N. A. Sinitsyn, Benchmarking Information Scrambling, \prl {\bf 129}, 050602 (2022).

\bibitem{Hayden07}
P. Hayden and J. Preskill, Black holes as mirrors: quantum information in random subsystems, J. High Energy Phys. {\bf 09}, 120 (2007).

\bibitem{Zanardi21}
P. Zanardi and N. Anand, Information scrambling and chaos in open quantum systems, \pra {\bf 103}, 062214 (2021).

\bibitem{Touil21}
A. Touil and S. Deffner, Information Scrambling versus Decoherence---Two Competing Sinks for Entropy, PRX QUANTUM {\bf 2}, 010306 (2021).

\bibitem{Prakash20}
R. Prakash and A. Lakshminarayan, Scrambling in strongly chaotic weakly coupled bipartite systems:
Universality beyond the Ehrenfest timescale, \prb {\bf 101}, 121108 (2020).

\bibitem{TianciZhou20}
T. Zhou, S. Xu, X. Chen, A. Guo, and B. Swingle, Operator L{\'e} vy Flight: Light Cones in Chaotic Long-Range Interacting Systems, \prl {\bf 124}, 180601 (2020).

\bibitem{SKZhao22}
S. K. Zhao, Z. Y. Ge, Z. Xiang, G. M. Xue, and S. P. Zhao, Probing Operator Spreading via Floquet Engineering in a Superconducting Circuit, \prl {\bf 129}, 160602 (2022).

\bibitem{Yin21}
C. Yin and A. Lucas, Quantum operator growth bounds for kicked tops and semiclassical spin chains, \pra {\bf 103}, 042414 (2021).

\bibitem{Moudgalya19}
S. Moudgalya, T. Devakul, C. V. Keyserlingk, and S. L. Sondhi, Operator spreading in quantum maps, \prb {\bf 99}, 094312 (2019).

\bibitem{RHFan17}
R. Fan, P. Zhang, H. Shen, and H. Zhai, Out-of-time-order correlation for many-body localization, Science Bulletin {\bf 62}(10):707-711 (2017).

\bibitem{Smith19}
A. Smith, J. Knolle, R. Moessner, and D. L. Kovrizhin, Logarithmic Spreading of Out-of-Time-Ordered Correlators without Many-Body Localization, \prl {\bf 123}, 086602 (2019).

\bibitem{Garttner18}
M. G{\"a}rttner, P. Hauke, and A. M. Rey, Relating Out-of-Time-Order Correlations to Entanglement via Multiple-Quantum Coherences, \prl {\bf 120}, 040402 (2018).

\bibitem{Keyserlingk18}
C. V. Keyserlingk, T. Rakovszky, F. Pollmann, and S. L. Sondhi, Operator Hydrodynamics, OTOCs, and Entanglement Growth in Systems without Conservation Laws, Phys. Rev. X {\bf 8}, 021013 (2018).

\bibitem{Lerose20}
A. Lerose and S. Pappalardi, Bridging entanglement dynamics and chaos in semiclassical systems, \pra {\bf 102}, 032404 (2020).

\bibitem{Lewis-Swan19}
R. J. Lewis-Swan, A. Safavi-Naini, J. J. Bollinger, and A. M.Rey, Unifying scrambling, thermalization and entanglement through measurement of fidelity out-of-time-order correlators in the Dicke model, Nat. Commun. {\bf 10}, 1581 (2019).

\bibitem{Zhu22}
Q. Zhu, Z. H. Sun, M. Gong, F. Chen, Y. R. Zhang, Y. Wu, Y. Ye, C. Zha, S. Li, and S. Guo, Observation of Thermalization and Information Scrambling in a Superconducting Quantum Processor, \prl {\bf 128}, 160502 (2022).

\bibitem{Balachandran21}
V. Balachandran, G. Benenti, G. Casati, and D. Poletti, From the eigenstate thermalization hypothesis to algebraic relaxation of OTOCs in systems with conserved quantities, \prb {\bf 104}, 104306 (2021).

\bibitem{Kobrin21}
B. Kobrin, Z. Yang, G. D. Kahanamoku-Meyer, C. T. Olund, J. E. Moore, D. Stanford, and N. Y. Yao, Many-Body Chaos in the Sachdev-Ye-Kitaev Model, \prl {\bf 126}, 030602 (2021).

\bibitem{Borgonovi19}
F. Borgonovi, F. M. Izrailev, and L. F. Santos, Timescales in the quench dynamics of many-body quantum systems: Participation ratio versus out-of-time ordered correlator, \pre {\bf 99}, 052143 (2019).


\bibitem{Shenglong22}
S. Xu and B. Swingle, Scrambling Dynamics and Out-of-Time Ordered Correlators in Quantum Many-Body Systems: a Tutorial, arXiv:2202.07060 quant-ph.

\bibitem{Heyl2018}
 M. Heyl, F. Pollmann, and B. D\'ora, Detecting equilibrium and dynamical quantum phase transitions in Ising chains via out-of-time-ordered correlators, \prl {\bf 121}, 016801 (2018).

\bibitem{Mi21}
X. Mi, {\it et al}., Information scrambling in quantum circuits, Science {\bf 374}, 1479 (2021).

\bibitem{Nie20}
X. Nie, {\it et al}., Experimental Observation of Equilibrium and Dynamical Quantum Phase Transitions via Out-of-Time-Ordered Correlators, \prl {\bf 124}, 250601 (2020).

\bibitem{LJZhai20}
L. J. Zhai and S. Yin, Out-of-time-ordered correlator in non-Hermitian quantum systems, \prb {\bf 102}, 054303 (2020).

\bibitem{WlZhao22}
W. L. Zhao, Quantization of out-of-time-ordered correlators in non-Hermitian chaotic systems, Phys. Rev. Research {\bf 4}, 023004 (2022).

\bibitem{Berry2004}
M. Berry, Physics of Nonhermitian Degeneracies, Czech. J. Phys. {\bf 54}, 1039 (2004).

\bibitem{Ashida20}
Y. Ashida, Z. Gong, and M. Ueda, Non-Hermitian physics, Adv. Phys. {\bf 69}, 249 (2020).

\bibitem{Bender1998}
C. M. Bender and S. Boettcher, Real Spectra in Non-Hermitian Hamiltonians Having PT Symmetry, \prl {\bf 80}, 5243 (1998).

\bibitem{Bender2002}
C. M. Bender, D. C. Brody, and H. F. Jones, Complex Extension of Quantum Mechanics, \prl {\bf 89}, 270401 (2002).


\bibitem{Zhaoxm2021}
X. M. Zhao, C. X. Guo, M. L. Yang, H. Wang, W. M. Liu, and S. P. Kou, Anomalous non-Abelian statistics for non-Hermitian generalization of Majorana zero modes, \prb {\bf 104}, 214502 (2021).

\bibitem{Yu2021}
Z. F. Yu, J. K. Xue, L. Zhuang, J. Zhao, and W. M. Liu, Non-Hermitian spectrum and multistability in exciton-polariton condensates, \prb {\bf 104}, 235408 (2021).

\bibitem{Zhaoxm20211}
X. M. Zhao, C. X. Guo, S. P. Kou, L. Zhuang, and W. M. Liu, Defective Majorana zero modes in a non-Hermitian Kitaev chain, \prb {\bf 104}, 205131 (2021).

\bibitem{Javier22}
J. D. Pino, J. J. Slim, and E. Verhagen, Non-Hermitian chiral phononics through optomechanically induced squeezing, Nature (London) {\bf 606}, 82 (2022).

\bibitem{Ganainy18}
R. El-Ganainy, K. G. Makris, M. Khajavikhan, Z. H. Musslimani, S. Rotter, and D. N. Christodoulides, Non-Hermitian physics and PT symmetry, Nat. Phys. {\bf 14}, 11 (2018).

\bibitem{Xiao19}
Z. Xiao, H. Li, T. Kottos, and A. Al{\ u}, Enhanced Sensing and Nondegraded Thermal Noise Performance Based on PT-Symmetric Electronic Circuits with a Sixth-Order Exceptional Point, \prl {\bf 123}, 213901 (2019).

\bibitem{Chitsazi17}
M. Chitsazi, H. Li, F. M. Ellis, and T. Kottos, Experimental Realization of Floquet $\cal{PT}$-Symmetric Systems, \prl {\bf 119}, 093901 (2017).

\bibitem{Deyuan21}
D. Zou, T. Chen, W. He, J. Bao, C. H. Lee, H. Sun, and X, Zhang, Observation of hybrid higher-order skin-topological effect in non-Hermitian topolectrical circuits, Nat. Commun. {\bf 12} 7201 (2021).

\bibitem{Kreibich16}
M. Kreibich, J. Main, H. Cartarius, and G. Wunner, Tilted optical lattices with defects as realizations of PT symmetry
in Bose-Einstein condensates, \pra {\bf 93}, 023624 (2016).

\bibitem{Keller97}
C. Keller, M. K. Oberthaler, R. Abfalterer, S. Bernet, J. Schmiedmayer, and A. Zeilinger, Tailored Complex Potentials and Friedel's Law in Atom Optics, \prl {\bf 79}, 3327 (1997).

\bibitem{Li19}
J. Li, A. K. Harter, J. Liu, L. D. Melo, Y. N. Joglekar, and L. Luo, Observation of parity-time symmetry breaking transitions
in a dissipative Floquet system of ultracold atoms, Nat. Commun. {\bf 10}, 855 (2019).

\bibitem{YongmeiXue}
Y. Xue, C. Hang, Y. He, Z. Bai, Y. Jiao, G. Huang, J. Zhao, and S. Jia, Experimental observation of partial parity-time symmetry and its phase transition with a laser-driven cesium atomic gas, \pra {\bf 105}, 053516 (2022).

\bibitem{Zejian22}
Z. Ren, D. Liu, E. Zhao, C. He, K. K. Pak, J. Li, and Gyu-Boong Jo, Chiral control of quantum states in non-Hermitian spin–orbit-coupled fermions, Nat. Phys. {\bf 18}, 385 (2022).

\bibitem{Longwen22}
L. Zhou and W. Han, Driving-induced multiple $\cal{PT}$-symmetry breaking transitions and reentrant localization transitions in non-Hermitian Floquet quasicrystals, Phys. Rev. B {\bf 106}, 054307 (2022).

\bibitem{Longwen21}
L. Zhou, Floquet engineering of topological localization transitions and mobility edges in one-dimensional non-Hermitian quasicrystals. Phys. Rev. Research {\bf 3}, 033184 (2021).

\bibitem{Longwen21B}
L. Zhou, Y. Gu, and J. Gong, Dual topological characterization of non-Hermitian Floquet phases, Phys. Rev. B {\bf 103}, L041404 (2021).

\bibitem{DaJian19}
D. J. Zhang, Q. Wang, and J. Gong, Time-dependent $\cal{PT}$-symmetric quantum mechanics in generic non-Hermitian systems, Phys. Rev. A {\bf 100}, 062121 (2019).

\bibitem{Longhi17}
S. Longhi, Oscillating potential well in the complex plane and the adiabatic theorem, Phys. Rev. A {\bf 96}, 042101 (2017).

\bibitem{West10}
C. T. West, T. Kottos, and T. Prosen, $\cal{PT}$ -Symmetric Wave Chaos, Phys. Rev. Lett. {\bf 104}, 054102 (2010).

\bibitem{Longhi2017}
S. Longhi, Localization, quantum resonances, and ratchet acceleration in a periodically kicked PT -symmetric quantum rotator, \pra {\bf 95} 012125 (2017).

\bibitem{Zhao19}
W. L. Zhao, J. Wang, X. Wang, and P. Tong, Directed momentum current induced by the PT-symmetric driving, \pre {\bf 99}, 042201 (2019).

\bibitem{Santhanam22}
M. S. Santhanam, S. Paul, and J. B. Kannan, Quantum kicked rotor and its variants: Chaos, localization and beyond Phys. Rep. {\bf 956}, 1 (2022).

\bibitem{Ho12}
D. Y. H. Ho, and J. Gong, Quantized Adiabatic Transport inMomentum Space, Phys. Rev. Lett. {\bf 109}, 010601 (2012).

\bibitem{Gadway13}
B. Gadway, J. Reeves, L. Krinner, and D. Schneble, Evidence for a Quantum-to-Classical Transition in a Pair of Coupled Quantum Rotors, Phys. Rev. Lett. {\bf 110}, 190401 (2013).

\bibitem{Huang21}
K. Q. Huang, W. L. Zhao, and Z. Li, Effective protection of quantum coherence by a non-Hermitian driving potential, Phys. Rev. A {\bf 104}, 052405 (2021).

\bibitem{Vuatelet21}
V. Vuatelet, and A. Ran{\c{c}}on, Effective thermalization of a many-body dynamically localized Bose gas, Phys. Rev. A {\bf 104}, 043302 (2021).

\bibitem{Belyansky20}
R. Belyansky, P. Bienias, Y. A. Kharkov, A. V. Gorshkov , and B. Swingle, Minimal Model for Fast Scrambling, \prl 125, 130601 (2020).

\bibitem{Kuwahara21}
T. Kuwahara, and K. Saito, Absence of Fast Scrambling in Thermodynamically Stable Long-Range
Interacting Systems, \prl {\bf 126}, 030604 (2021).


\bibitem{DAlessio14}
L. D'Alessio and M. Rigol, Long-time Behavior of Isolated Periodically Driven Interacting Lattice Systems, Phys. Rev. X
{\bf 4}, 041048 (2014).




\bibitem{Zhao21}
W. L. Zhao, Y. Hu, Z. Li, and Q. Wang, Super-exponential growth of out-of-time-ordered correlators, \prb {\bf 103}, 184311 (2021).


\bibitem{Huang19}
Y. Huang, F. G. S. L. Brand{\~a}o, and Y. L. Zhang, Finite-Size Scaling of Out-of-Time-Ordered Correlators at Late Times, \prl {\bf 123}, 010601 (2019).

\bibitem{Wei19}
B. B. Wei, G. Sun, and M. J. Hwang, Dynamical scaling laws of out-of-time-ordered correlators, \prb {\bf 100}, 195107 (2019).


\end{thebibliography}
\end{document}